\documentclass[pra,aps,reprint,showpacs,amsmath]{revtex4-1}

\usepackage{bm}
\usepackage{graphicx}
\usepackage{dsfont}

\newcommand{\bra}[1]{\langle #1 | \,}
\newcommand{\ket}[1]{\, | #1 \rangle}

\newcommand{\expv}[1]{\langle #1 \rangle}

\newcommand{\Om}{\Omega}
\newcommand{\ga}{\gamma}
\newcommand{\Ga}{\Gamma}
\newcommand{\de}{\delta}
\newcommand{\De}{\Delta}

\newcommand{\la}{\lambda}

\newcommand{\db}{d_{\mathrm{b}}}
\newcommand{\nR}{n_{\mathrm{R}}}
\newcommand{\pR}{p_{\mathrm{R}}}
\newcommand{\hlf}{\frac{1}{2}}

\newcommand{\sig}{\hat{\sigma}}

\newcommand{\sigav}{\bar{\sigma}}
\newcommand{\hS}{\hat{S}}

\begin{document}

\title{Two-dimensional crystals of Rydberg excitations 
in a resonantly driven lattice gas}

\author{David Petrosyan}
\affiliation{Institute of Electronic Structure and Laser, 
FORTH, GR-71110 Heraklion, Crete, Greece}
\date{\today}

\begin{abstract}
The competition between resonant optical excitation of Rydberg 
states of atoms and their strong, long-range van der Waals interaction
results in spatial ordering of Rydberg excitations in a 
two-dimensional lattice gas, as observed in a recent experiment 
of Schau\ss \ \textit{et al.} [Nature \textbf{491}, 87 (2012)].
Here we use semiclassical Monte Carlo simulations to obtain 
stationary states for hundreds of atoms in finite-size lattices. 
We show the formation of regular spatial structures of 
Rydberg excitations in a system of increasing size, and find 
highly sub-Poissonian distribution of the number of Rydberg 
excitations characterized by a large negative value of the 
Mandel $Q$ parameter which is nearly independent of the system size.
\end{abstract}

\pacs{37.10.Jk, 
32.80.Ee, 
75.30.Fv  
}

\maketitle

\section{Introduction}

Cold atoms in optical lattices represent a remarkably clean and 
controllable systems to simulate and study many-body physics \cite{OptLatRev}. 
A crucial ingredient for realizing various phases of matter is 
availability of interactions with different strength and range. 
The interaction between the ground-state atoms in a lattice is typically 
short range (on-site) and can be tuned relative to the (inter-site) 
tunneling energy to realize the Hubbard's model transition from
the superfluid to the Mott insulator phase with precisely one atom 
per lattice site \cite{OL-Hub}. 
Second order tunneling \cite{Trotzky20078} can translate into an effective 
nearest-neighbor interaction which allows the realization of the Heisenberg
spin or extended Hubbard models \cite{Kuklov2003,Petrosyan2007}. 
Dipolar interactions \cite{Lahaye2009} between atoms with large magnetic 
moment or between polar molecules have still longer range. 
But atoms excited to the Rydberg states exhibit unprecedented 
magnitude and range of dipole-dipole or van der Waals 
interactions \cite{RydAtoms,rydrev,Loew2012}. Dressing ground 
state atoms by the Rydberg state with an off-resonant laser field
can thus lead to an effective long-range interaction 
\cite{Johnson2010,Henkel2010,Pupillo2010,Lauer2012}. 

An alternative approach is to explore resonant coupling of the ground 
state atoms to the highly excited Rydberg states \cite{Loew2012}. 
The interaction between the atoms in Rydberg states translates into
the level shifts of multiple Rydberg excitations which are therefore 
suppressed in dense atomic ensembles
\cite{Tong2004,Singer2004,Vogt2006,Heidemann2007,Low2009,RdSV2013}. 
Thus, within a small volume, where the interatomic interaction energies 
exceed the excitation linewidth of the Rydberg state, a single 
Rydberg atom blocks the excitation of all the other atoms 
\cite{Lukin2001,Robicheaux2005,Stanojevic2009,UrbanGaetan2009,%
Dudin2012NatPh,DPKM2013,Ates2013}. Larger atomic ensembles 
can accommodate more Rydberg excitations which effectively repel 
each other so that no two Rydberg atoms can appear closer than the 
blockade distance \cite{Schwarzkopf2011,Viteau2011,Schauss2012}.
Simultaneously, the total number of excitations within the ensemble
exhibits reduced fluctuations \cite{Raithel2005,Viteau2012,Hofmann2013}. 

Spatial correlations and crystallization of Rydberg excitations
have recently invigorated many theoretical investigations 
\cite{Pupillo2010,Lauer2012,Weimer2008,Weimer10,Pohl2010,Schachenmayer2010,%
Bijnen2011,Sela2011,Lesanovsky2011,Lesanovsky2012,Zeller2012,Garttner2012,%
Lee2011,Qian2012,Hoening2013,Ji2011,AtesGarraha2012,Ates2012Ji2013,DPMHMF2013}.
A salient experiment of Schau\ss \ \textit{et al.} \cite{Schauss2012}
have demonstrated spatial ordering of Rydberg excitations 
of resonantly driven atoms in a two-dimensional (2D) optical lattice. 
In an atomic cloud with the diameter of about two blockade distances, 
up to five Rydberg excitations were observed. 
In that experiment, the lifetime of the Rydberg state and the coherence time 
of the atomic polarization were much longer than the duration ($\sim \mu$s) 
of the optical excitation pulses. Hence, neglecting the relaxations, 
numerical simulations of the coherent excitation dynamics of many-body system
well reproduced the experimental observations \cite{Schauss2012}. 

The purpose of the present study is to obtain and characterize the many-body 
stationary state of a strongly interacting dissipative system subject 
to continuous (long-time) excitation. To that end, we employ an iterative 
Monte Carlo sampling algorithm \cite{DPMHMF2013} to simulate the steady-state
distribution of Rydberg excitations of the van der Waals interacting atoms 
in a 2D lattice of variable size. As shown in \cite{Petrosyan2013}, however, 
the state of the system resulting after only a few $\mu$s driving is 
already close to the steady state. Similarly to \cite{Schauss2012}, we obtain 
spatially ordered patterns formed by definite numbers of Rydberg excitations, 
while the probability distribution of the excitation numbers is 
very narrow (highly sub-Poissonian) characterized by a large negative 
value of the Mandel $Q$ parameter \cite{Raithel2005,Viteau2012,Hofmann2013}.
In a moderately sized atomic cloud, a few mutually repelling Rydberg 
excitations dwell near the circular boundary of the system, as was also 
observed in \cite{Schauss2012}. With increasing the size (diameter $d$) 
of the cloud, while keeping the atomic density fixed, the mean number 
of Rydberg excitations grows ($\propto d^{5/3}$), but its distribution 
remains sub-Poissonian with nearly constant $Q \approx -0.84$. 
Concurrently, the Rydberg excitations tend to arrange in regular spatial 
shell structures, which are, however, progressively smeared for larger 
number of excitations. The corresponding density-density correlations 
decay on the length scale of the blockade distance.

The paper is structured as follows: After describing the Monte Carlo 
procedure in Sec.~\ref{sec:MC}, we present the results of numerical
simulations in Sec.~\ref{sec:Results}, followed by the discussion 
and conclusions in Sec.~\ref{sec:Discuss}.

\section{The Monte Carlo algorithm}
\label{sec:MC}

Strongly interacting many-body systems are rarely amenable to exact 
numerical treatment due to the exponential scaling of the corresponding
Hilbert space with the system size. For an ensemble of $N$ two-level 
atoms interacting via a long-range potential, truncating the Hilbert space 
by limiting the number of Rydberg excitations can greatly reduce the 
dimension of the problem and allow approximate solutions for tens 
or hundreds of atoms sharing several excitations
\cite{Schauss2012,Viteau2012,Garttner2012,Petrosyan2013}. 
As an alternative technique, semiclassical Monte Carlo simulations 
\cite{Hofmann2013,DPMHMF2013,Ates2007,Ates2011,Heeg2012} can 
efficiently deal with many more atoms and excitations. The essence 
of the semiclassical methods is to treat each atom separately 
in a potential generated by all the other atoms excited to 
the Rydberg state, while neglecting the inter-atomic coherences 
(entanglement). This approximation is valid when the dephasing
rate of the atomic polarization is larger than the Rabi 
frequency of the driving field \cite{DPMHMF2013}.

Let us recall the properties of a single two-level atom 
driven by a laser field with Rabi frequency $\Om$ and detuning
$\de$ on the transition from the ground state $\ket{g}$ to the 
excited (Rydberg) state $\ket{r}$. We denote by $\Ga_r$ the decay 
rate of the excited state population $\expv{\sig_{rr}}$, and by 
$\Ga_z$ the relaxation rate of atomic polarization $\expv{\sig_{rg}}$, 
where $\sig_{\mu \nu} \equiv \ket{\mu}\bra{\nu}$ is the projection 
($\mu = \nu$) or transition ($\mu \neq \nu$) operator of the atom. 
The steady-state population of the Rydberg state is then 
a Lorentzian function of detuning $\de$ \cite{PLDP2007},
\begin{equation}
\expv{\sig_{rr}} = \frac{\Om^2}
{2 \Om^2 + \frac{\Ga_r}{2\ga_{rg}} (\ga_{rg}^2 + \de^2) } ,
\label{sigrr}  
\end{equation}
with the width $w \simeq 2 \Om \sqrt{\ga_{rg}/\Ga_r}$,
where $\ga_{rg} \equiv \hlf \Ga_{r} + 2 \Ga_{z}$ and we assume strong 
driving $\Om^2 > \Ga_r \ga_{rg}$ (or weak decay $\Ga_r < \Om^2 /\ga_{rg}$).  
On resonance, $\de \ll w$, the population saturates to 
$\expv{\sig_{rr}} \to \hlf$.

Atoms in the Rydberg state $\ket{r}$ interact pairwise via the 
van der Waals (vdW) potential, $\hbar \De(z) = \hbar C_6/z^6$, where $C_6$ 
is the vdW coefficient and $z$ is the interatomic distance  \cite{rydcalc}. 
Hence, given an atom $i$ in state $\ket{r}$, it will induce a level 
shift $\De(z_{ij})$ of another atom $j$, which effectively 
translates into the detuning $\de$. When $\De(z_{ij}) \gtrsim w$,
the transition $\ket{g} \to \ket{r}$ of atom $j$ is non-resonant 
and its Rydberg excitation is blocked by atom $i$ \cite{Lukin2001,rydrev}.
We can therefore define the blockade distance $\db$ via
$\De(\db) = w$ which yields $\db = \sqrt[6]{C_6/w}$.

Our aim is to obtain the stationary distribution of Rydberg excitations
of $N \gg 1$ atoms in a 2D lattice. We make a semiclassical
approximation which discards quantum correlations between the atoms 
but preserves classical $N$-body correlations. Its validity hinges 
on the assumption of strong dephasing $\Ga_z \gtrsim \Om$ which 
suppresses intra- and interatomic coherences and disentangles the atoms
\cite{DPMHMF2013}. Each atom $j$ then behaves as a driven two-level 
system of Eq.~(\ref{sigrr}) but with the detuning $\de$ determined by 
operator $\hS_j \equiv \sum_{i \neq j}^N \sig_{rr}^i \Delta(z_{ij})$ which 
describes the total interaction-induced shift of its level $\ket{r}$ 
involving the contributions of all the other Rydberg atoms $\sig_{rr}^i$. 

We employ the procedure described in \cite{DPMHMF2013} to generate 
the stationary distribution of Rydberg excitations in an ensemble 
of $N \lesssim 10^3$ atoms. The algorithm performs an iterative 
Monte Carlo sampling of $\{ \sig_{rr}^j \}$ for $N$ atoms, 
in the spirit of the Hartree-Fock method. 
We first initialize all the atoms in, e.g., the ground state, 
$\sig_{gg}^j \to 1 \, \forall \, j \in [1,N]$. 
Then, at every step, for each atom $j$, we set $\sig_{rr}^j \to 1$ or $0$
with the probability determined by its current Rydberg state population 
$\expv{\sig_{rr}^j}$. In turn, the thus constructed binary configuration 
of Rydberg excitations $\{ \sig_{rr}^i \} \to \{ 0,1,0,0 \ldots \}$ 
determines the level shift $\hS_j$ (equivalent to detuning $\de$) of 
atom $j$ when evaluating $\expv{\sig_{rr}^j}$. We continuously iterate 
this procedure, sifting repeatedly through every atom in the potential 
generated by all the other atoms. We thus generate a large number 
($\sim 10^6$) of configurations $\{ \sig_{rr}^j \}$ from which we
obtain the averaged probability distribution $\sigav_{rr}^j$ of Rydberg 
excitations and their correlations $\overline{\sig_{rr}^i \sig_{rr}^j}$, 
as well as the probabilities $\pR(n)$ of $n$ Rydberg excitations and
their spatial patterns. 

Note that in the experiment, after preparing the atoms in the ground 
state $\ket{g}$ and exciting them with a laser, one performs projective 
measurements $\{ \sig_{rr}^j \}$ of Rydberg excitations \cite{Schauss2012}.
Every experimental sequence then results in a particular configuration 
$\{ \sig_{rr}^j \} \to \{1,0,0,1,\ldots \}$, and averaging over many such
experimental sequences yields the Rydberg excitation probabilities for 
individual atoms or the whole ensemble. Hence, the Monte Carlo sampling 
of the excitation configurations closely imitates the experiment.

\section{Results of numerical simulations}
\label{sec:Results}

In our simulations, we use the parameters similar to those in the 
experiment \cite{Schauss2012}, with the exception of a larger relaxation 
rate $\Ga_z$ compatible with the semiclassical approximation.
We thus assume a 2D lattice of cold $^{87}$Rb atoms whose ground 
state $\ket{g} \equiv 5 S_{1/2} \ket{F=2,m_F=-2}$ is coupled  
to the Rydberg state $\ket{r} \equiv 43 S_{1/2}$ by a two-photon
transition with the Rabi frequency $\Om/2\pi = 100\:$kHz.
The decay rate of $\ket{r}$ is $\Ga_r = 0.065 \Om \simeq 40\:$kHz and 
we take large enough coherence relaxation rate $\Ga_z = \Om \simeq 630\:$kHz
(in the experiment $\Ga_z \simeq 160\:$kHz, stemming from the two-photon
laser linewidth $\sim 70\:$kHz and residual decay $\sim 90\:$kHz of 
the atomic intermediate state to the ground state \cite{Schauss2012}). 
The resulting excitation linewidth of $\ket{r}$ is 
$w / 2\pi \simeq 0.8\:$MHz. With the vdW coefficient
$C_6/ 2\pi \simeq 2.45 \: \mathrm{GHz} \: \mu\mathrm{m}^6$ \cite{rydcalc}
the corresponding blockade distance is $\db \simeq 3.81\:\mu$m. 
We performed numerical simulations for various diameters $d$ of 
the circular boundaries enclosing the 2D lattice of atoms at constant 
density (fixed lattice spacing $a= 532\:$nm \cite{Schauss2012}).  

\begin{figure*}[t]
\includegraphics[width=18.0cm]{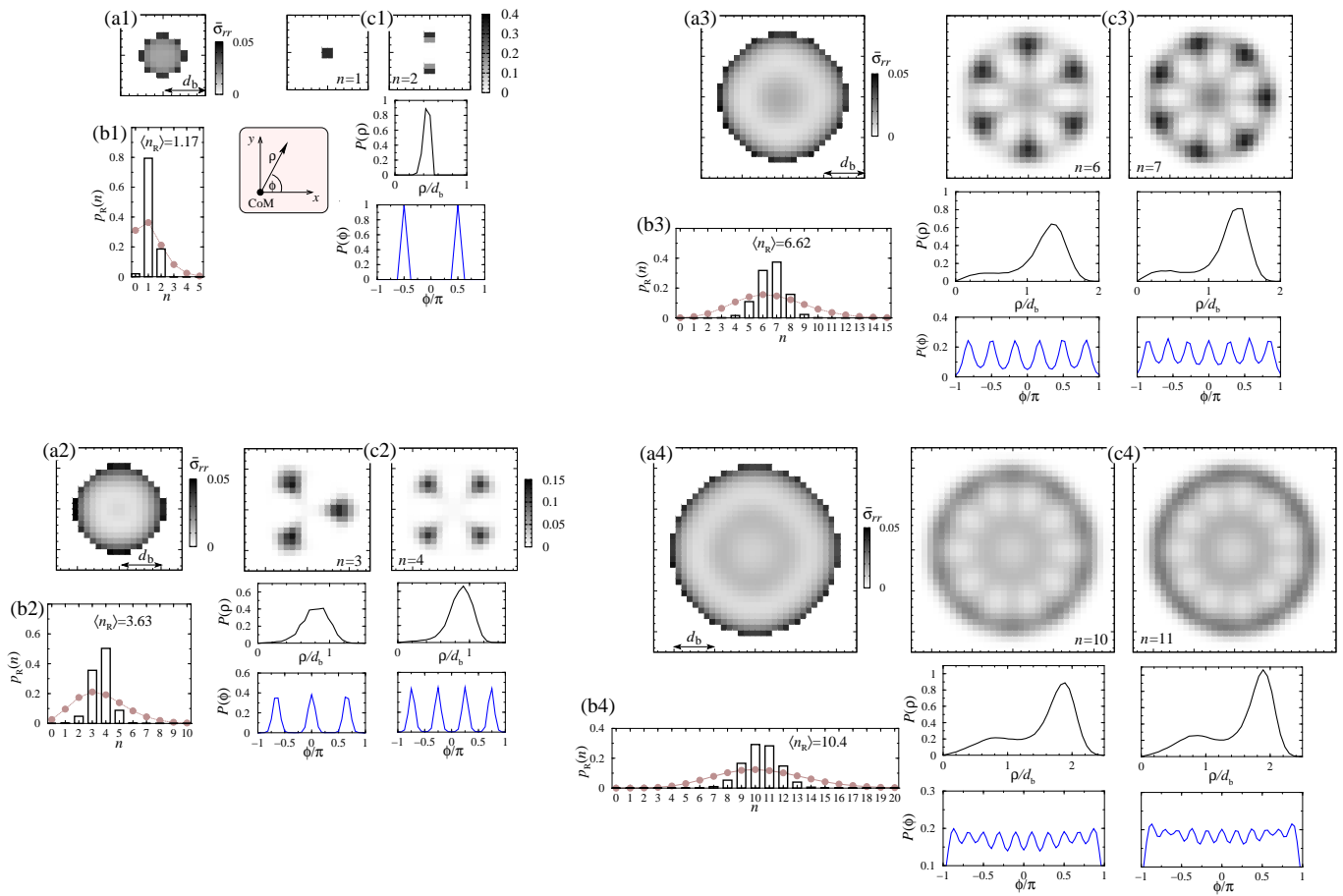}
\caption{(Color online)
Rydberg excitations of a resonantly-driven ensemble of 
$N = (44,188,401,688)$ atoms in a 2D lattice of diameter 
$d = (1,2,3,4)\db$, respectively.
(a1-a4)~Average spatial distribution of Rydberg excitation 
probabilities $\sigav_{rr}$; each pixel corresponds to a single 
atom/lattice site. 
(b1-b4)~Probabilities $\pR(n)$ of $n$ Rydberg excitations (open bars) 
and the mean number of excitations $\expv{\nR}$. 
Also shown is the Poisson distribution $p_{\mathrm{Poisson}}(n)$ 
for the same $\expv{\nR}$ (solid brown circles).
(c1-c4)~Average spatial distributions of $n$ Rydberg excitations, 
corresponding to two largest $\pR(n)$ in (b1-b4), after centering 
and aligning each configuration $\{ \sig_{rr}^j \}_n$, and the 
corresponding axial $P(\rho)$ and angular $P(\phi)$ probability 
distributions [$P(\rho)$ are binned over the lattice spacing $a$  
and $P(\phi)$ are binned over the interval $\pi a/d$].} 
\label{fig:L1234c}
\end{figure*}

In Fig.~\ref{fig:L1234c} we present the results of simulations for 
$d =(1,2,3,4)\db$. Part (a) of each panel shows the Rydberg 
excitation probabilities $\sigav_{rr}$ averaged over all the
configurations for a given diameter $d$ of the atomic cloud. 
The spatial distribution of Rydberg excitations is 
rotationally symmetric, with the largest probabilities 
at the boundary of the cloud. With increasing the system size, 
however, we observe a formation of spatial shell structure
of Rydberg excitations [the inner ring in Fig.~\ref{fig:L1234c}(a4)].

\begin{figure}[t]
\centerline{\includegraphics[width=6cm]{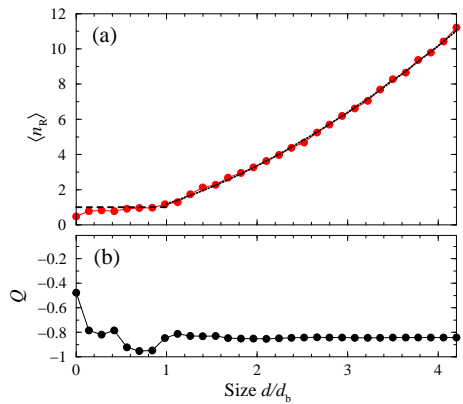}}
\caption{(Color online)
(a) Mean number of Rydberg excitations $\expv{\nR}$, and 
(b) the corresponding $Q$ parameter vs the system size $d/\db$.
In (a) the the black dashed line for $d < \db$ is $\expv{\nR} =1$,
and the dotted line for $d \geq \db$ is $\expv{\nR} = 0.2 + (d/\db)^{1.66}$.}
\label{fig:nRQ}
\end{figure}

In Figs.~\ref{fig:L1234c}(b1-b4), we show the mean number of 
Rydberg excitations in the cloud $\expv{\nR} = \sum_n n \pR(n)$ 
and the probabilities $\pR(n)$ to find $n=0,1,2,\ldots$ excitations.
Interestingly, $\expv{\nR}$ grows with the size of the system 
as $\expv{\nR} \propto (d/\db)^{5/3}$ [Fig.~\ref{fig:nRQ}(a)], 
rather than $(d/\db)^2$ as one might expect for a 2D bulk system. 
This is a finite-size effect, due to the larger concentration of 
the mutually repelling Rydberg excitations at the cloud boundary
whose circumference grows linearly with $(d/\db)$. 
[We may conjecture that in a 3D system $\expv{\nR} \propto (d/\db)^{\alpha}$
with $2 < \alpha < 3$ due to the concentration of Rydberg excitations 
on the surface of the sphere.]

To quantify the number distribution of Rydberg excitations, 
we use the Mandel $Q$ parameter \cite{MandelQ}
\begin{equation}
Q \equiv \frac{\expv{\nR^2} - \expv{\nR}^2}{\expv{\nR}} - 1 ,
\end{equation}
where $\expv{\nR^2} = \sum_n n^2 \, \pR(n)$. A Poissonian 
distribution $p_{\mathrm{Poisson}}(n) =  \expv{\nR}^n e^{-\expv{\nR}}/n!$
would lead to $Q=0$, while $Q < 0$ corresponds to sub-Poissonian
distribution, with $Q=-1$ attained for a definite number $n$ 
of excitations. We find highly sub-Poissonian distribution
of $\pR(n)$, Figs.~\ref{fig:L1234c}(b1-b4), and nearly constant
$Q \simeq -0.84 \; \forall \; d > \db$, Fig.~\ref{fig:nRQ}(b).
Note that in the absence of coherence realization, $\Ga_z =0$, 
the (classical) correlations of the Rydberg excitations are considerably
smaller, $Q \sim -0.5$ \cite{Petrosyan2013} (see Sec.~\ref{sec:Discuss}).

Next, for each $d$, we select the values of $n$ with the largest probabilities
$\pR(n)$ (frequent occurrence) and analyze the average spatial distribution
of the corresponding configurations $\{ \sig_{rr}^j \}_n$ containing
precisely $n$ Rydberg excitations, as was done in \cite{Schauss2012}. 
To this end, for each configuration $\{ \sig_{rr}^j \}_n$ we set 
the origin of the polar coordinate system at the center of mass (CoM)
of the excitations. We then determine the mean angle of the position 
vectors of all the excited atoms with respect to a reference ($x$) 
axis and rotate the configuration about the origin by that angle. 
The averages over many thus centered and aligned configurations are 
shown in Figs.~\ref{fig:L1234c}(c1-c4). Below each 2D density plot
(apart from that for $n=1$) we also show the corresponding axial 
$P(\rho)$ and angular $P(\phi)$ probability distributions of the 
$n$ excitations. We observe that the Rydberg excitations tend to 
arrange in regular spatial patterns, i.e., they crystallize.  
In a cloud of diameter $d \lesssim 2\db$, the $n$ excitations are pushed
to the boundary, and their angular separation is $\de \phi \sim 2\pi/n$.  
But already for $d \sim 3\db$ we encounter a significant probability 
of Rydberg excitation at the cloud center, while for $d \sim 4\db$
we clearly observe a double peak structure of $P(\rho)$, with the largest
peak close to the boundary of the cloud and the smaller peak corresponding
to the inside ring of the corresponding 2D density plot. The distance 
between the two peaks of $P(\rho)$ is somewhat larger than the blockade 
distance $\db$ (see below). Simultaneously, we observe reduced contrast
of the angular distribution $P(\phi)$. 

\begin{figure}[t]
\centerline{\includegraphics[width=6cm]{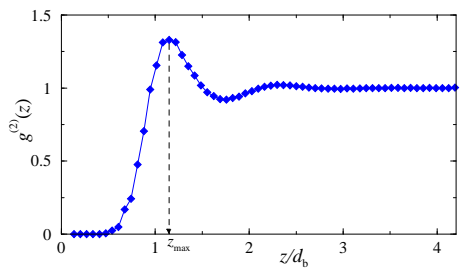}}
\caption{(Color online)
Correlations of Rydberg excitations $g^{(2)}(z)$ vs distance $z$;
the maximum is at $z_{\mathrm{max}} = 1.15\db$.}
\label{fig:g2}
\end{figure}

In Fig.~\ref{fig:g2} we plot the normalized spatial correlations 
\begin{equation}
g^{(2)}(z) \equiv 
\frac{\sum_{i \neq j} \de_{z,z_{ij}} \overline{\sig_{rr}^i \sig_{rr}^j}}
{\sum_{i \neq j} \de_{z,z_{ij}} \sigav_{rr}^i \sigav_{rr}^j} \, ,
\end{equation}
where $\de_{z,z_{ij}}$ is the Kronecker delta which selects the pairs
of atoms $i,j$ at distance $z_{ij}=z$. Clearly, at short distances
$z < \db$ the Rydberg excitations avoid each other 
\cite{Schwarzkopf2011,Viteau2011,Schauss2012}, while at $z_{\mathrm{max}}$
slightly larger than $\db$ there is a pronounced maximum of $g^{(2)}(z)$,
which implies higher probability of finding pairs of Rydberg 
excitations separated by distance $z_{\mathrm{max}}$. Precisely 
this suppression, or blockade, of Rydberg excitations at shorter 
distances, and the elevated probability of excitations at distance 
$z_{\mathrm{max}}$, are responsible for the observed quasi-crystallization 
and inner shell structure of the spatial distribution of $\sigav_{rr}^j$.
At longer distances $z > \db$, the correlation function $g^{(2)}(z)$ 
exhibits strongly damped spatial oscillations, signifying 
the absence of long-range order. Similar results were obtained 
for a 1D system in \cite{DPMHMF2013}, where it was found that the 
correlation length $\xi$ exceeding the oscillation period $\la \gtrsim \db$
can only be attained at very high atomic densities 
(thirty or more atoms per blockade distance $\db$).

\section{Discussion}
\label{sec:Discuss}

Above we have analyzed the steady-state distribution of Rydberg excitations
in a 2D cloud of atoms subject to continuous resonant driving. 
We have observed very narrow (sub-Poissonian) distribution of 
the number of excitations and the formation of regular spatial 
structures of definite numbers of Rydberg excitations. 
These quasi-crystals are effected by the boundary of the system and 
the effective repulsion between the Rydberg excitations which avoid 
each other at distances smaller than the blockade distance $\db$. 
A cloud of diameter $d \lesssim 2\db$ can then accommodate only 
a few Rydberg excitations which dwell near the boundary.
With increasing the cloud size $d$, the Rydberg quasi-crystals grow, 
forming spatial shell structures, but their contrast decreases and beyond 
$d \gtrsim 4\db$ they are hardly discernible. The absence of the long-range
order in Rydberg quasi-crystals, also attested by the rapid decay of the
density-density correlation function $g^{(2)}(z)$, is due to the ``softness'' 
of the van der Waals potential and the finite excitation linewidth $w$ 
of the Rydberg state. Yet, the Mandel $Q$ parameter, which quantifies 
the counting statistics of the Rydberg excitations, remains negative
and nearly independent of the system size.

Sub-Poissonian statistics of the number of Rydberg excitations and 
the corresponding negative values of the Mandel $Q$ parameter were 
observed in several experiment \cite{Raithel2005,Viteau2012,Hofmann2013}
which dealt with the transient unitary dynamics of the system. 
It is interesting to note that for a coherently driven system 
$Q \sim - 0.5$ \cite{Petrosyan2013} while incoherent excitation 
studied here results in $Q < - 0.5$ and the Rydberg quasi-crystal 
exhibits sharper contrast than in \cite{Schauss2012}. In other words, 
dephasing leads to stronger classical (density-density) correlations 
of Rydberg excitation. This effect is inherited \cite{DPMHMF2013,Petrosyan2013}
from a single superatom---collection of atoms within the blockade distance: 
Without dephasing, the Rydberg population of a superatom saturates
to $\expv{\nR} \to \hlf$ ($Q=-0.5$); with strong dephasing and many atoms,
$\expv{\nR}$ approaches 1 ($Q=-1$), as seen in Fig.~\ref{fig:nRQ} for $d<\db$. 

In our simulations, we assumed an experimental setup similar to that
in \cite{Schauss2012}, i.e., square lattice with circular boundary,
obtaining, for a moderately sized atomic cloud, qualitatively 
similar results. Taking, instead, a different lattice geometry would 
lead to different spatial ordering of Rydberg excitations, but the more
fundamental quantities, $Q$ and $g^{(2)}$, would behave invariantly 
for a large cloud of the same atom density.

Let us finally discuss the limitations of our approach and the obtained
results. The two key questions are whether the steady state can be attained
in a real cold-atom experiment, especially for a large ensemble of 
$N \sim 10^3$ atoms, and to what extend the employed semiclassical 
treatment is adequate to describe the true steady state of the system.
These questions were addressed in \cite{Petrosyan2013}, 
at least for a moderate number of atoms $N \lesssim 50$, 
and here we briefly summarize the conclusions.
In an experiment with typical relaxation rates of the atoms \cite{Schauss2012},
the time scale for attaining the true many-body steady state is tens 
of $\mu$s---or more for larger $N$---which can be prohibitively long 
due to the loss of atoms and the need of continuous laser irradiation. 
However, already after several $\mu$s of laser driving, the spatial 
probability distribution of Rydberg excitations ceases to change 
appreciably \cite{Schauss2012} being close to the steady-state 
distribution \cite{Petrosyan2013}. Moreover, stronger coherence relaxation
$\Ga_z \gtrsim \Om$ will quickly dephase the atoms rendering the system 
essentially classical but still strongly correlated \cite{Petrosyan2013}.
The results of the semiclassical Monte Carlo simulations are then trustworthy, 
as was shown in \cite{DPMHMF2013}.


\begin{thebibliography}{99}

\bibitem{OptLatRev}
I.~Bloch, J.~Dalibard, and W.~Zwerger,
Rev. Mod. Phys. {\bf 80}, 885 (2008);
M.~Lewenstein, A.~Sanpera, V.~Ahufinger, B.~Damski, A.~Sen De, and U.~Sen,
Adv. Phys. {\bf 56},  243 (2007).

\bibitem{OL-Hub} 
D. Jaksch, C.  Bruder, J.I. Cirac, C.W. Gardiner, and P.~Zoller,
Phys. Rev. Lett. {\bf 81}, 3108 (1998);
M. Greiner, O. Mandel, T. Esslinger, T.W. H\"ansch, and I.~Bloch,
Nature {\bf 415}, 39 (2002);
D.~Jaksch and P.~Zoller,
Ann. Phys. (N.Y.) {\bf 315}, 52 (2005).

\bibitem{Trotzky20078}
S. F\"olling, S. Trotzky, P. Cheinet, M. Feld, R. Saers, A. Widera,
T. M\"uller, and I. Bloch,
Nature {\bf 448}, 1029 (2007);
S. Trotzky, P. Cheinet, S. F\"olling, M. Feld, U. Schnorrberger,
A.M. Rey, A. Polkovnikov, E.A. Demler, M.D. Lukin, and I. Bloch,
Science {\bf 319}, 295 (2008).

\bibitem{Kuklov2003}
A.B. Kuklov and B.V. Svistunov,
Phys. Rev. Lett. {\bf 90}, 100401 (2003);
L.-M. Duan, E. Demler, and M.D. Lukin,
Phys. Rev. Lett. {\bf 91}, 090402 (2003).

\bibitem{Petrosyan2007}
D. Petrosyan, B. Schmidt, J. R. Anglin and M. Fleischhauer, 
Phys. Rev. A {\bf 76}, 033606 (2007);
B. Schmidt, M. Bortz, S. Eggert, M. Fleischhauer and D. Petrosyan, 
Phys. Rev. A {\bf 79}, 063634 (2009).

\bibitem{Lahaye2009}
T. Lahaye, C. Menotti, L. Santos, M. Lewenstein, and T. Pfau,
Rep. Prog. Phys. {\bf 72} 126401 (2009).

\bibitem{RydAtoms}
T.F.~Gallagher, {\em Rydberg Atoms} (Cambridge University Press, 
Cambridge, 1994).

\bibitem{rydrev}
M. Saffman, T.G. Walker, and K. M\o lmer,
Rev. Mod. Phys. {\bf 82}, 2313 (2010);
D. Comparat and P. Pillet,
J. Opt. Soc. Am. B {\bf 27}, A208 (2010).

\bibitem{Loew2012}
R.  L\"ow, H. Weimer, J. Nipper, J.B. Balewski, B. Butscher, 
H.P. B\"uchler, and T. Pfau,
J. Phys. B {\bf 45}, 113001 (2012).

\bibitem{Johnson2010}
J.E. Johnson and S.L. Rolston, 
Phys. Rev. A {\bf 82}, 033412 (2010).

\bibitem{Henkel2010}
N. Henkel, R. Nath, and T. Pohl,
Phys. Rev. Lett. {\bf 104}, 195302 (2010).

\bibitem{Pupillo2010}
G. Pupillo, 
A. Micheli, M. Boninsegni, I. Lesanovsky, and P. Zoller,
Phys. Rev. Lett. {\bf 104}, 223002 (2010).

\bibitem{Lauer2012}
A. Lauer, D. Muth and M. Fleischhauer,
New J. Phys. {\bf 14} 095009 (2012).

\bibitem{Tong2004}
D. Tong, S.M. Farooqi, J. Stanojevic, S. Krishnan, Y.P. Zhang, R. C\^ot\'e, 
E.E. Eyler, and P.L. Gould, 
Phys. Rev. Lett. {\bf 93}, 063001 (2004).

\bibitem{Vogt2006}
T. Vogt, M. Viteau, J. Zhao, A. Chotia, D. Comparat, and P. Pillet,
Phys. Rev. Lett. {\bf 97}, 083003 (2006).

\bibitem{Singer2004}
K. Singer, M. Reetz-Lamour, T. Amthor, L.G. Marcassa, and M. Weidem\"uller,
Phys. Rev. Lett. {\bf 93}, 163001 (2004).

\bibitem{Heidemann2007}
R. Heidemann, U. Raitzsch, V. Bendkowsky, B. Butscher, R. L\"ow, 
L. Santos, and T. Pfau,
Phys. Rev. Lett. {\bf 99}, 163601 (2007).

\bibitem{Low2009}
R. L\"ow, H. Weimer, U. Krohn, R. Heidemann, V. Bendkowsky, B. Butscher, 
H.P. B\"uchler, and T. Pfau,
Phys. Rev. A {\bf 80}, 033422 (2009).

\bibitem{RdSV2013}
M. Robert-de-Saint-Vincent, C.S. Hofmann, H. Schempp, G. G\"unter, 
S. Whitlock, and M. Weidem\"uller,
Phys. Rev. Lett. {\bf 110}, 045004 (2013).

\bibitem{Lukin2001}
M.D. Lukin, M. Fleischhauer, R. C\^ot\'e, L.M. Duan, D. Jaksch, 
J.I. Cirac, and P. Zoller, 
Phys. Rev. Lett. {\bf 87}, 037901 (2001).

\bibitem{Robicheaux2005}
F. Robicheaux and J.V. Hernandez, 
Phys. Rev. A {\bf 72}, 063403 (2005).

\bibitem{Stanojevic2009}
J. Stanojevic and R. C\^ot\'e, 
Phys. Rev. A {\bf 80}, 033418 (2009).

\bibitem{UrbanGaetan2009}
E. Urban, T.A. Johnson, T. Henage, L. Isenhower, D.D. Yavuz, 
T.G. Walker, and M. Saffman,
Nature Phys. {\bf 5}, 110 (2009);
A. Ga\"etan, Y. Miroshnychenko, T. Wilk, A. Chotia, M. Viteau, D. Comparat, 
P. Pillet, A. Browaeys, and P. Grangier,
{\it ibid} {\bf 5}, 115 (2009).

\bibitem{Dudin2012NatPh}
Y.O. Dudin, L. Li, F. Bariani, and A. Kuzmich,
Nature Phys. {\bf 8}, 790 (2012).

\bibitem{DPKM2013}
D. Petrosyan and K. M\o lmer, 
Phys. Rev. A {\bf 87}, 033416 (2013).

\bibitem{Ates2013}
C. Ates, I. Lesanovsky, C.S. Adams, and K.J. Weatherill,
Phys. Rev. Lett. {\bf 110}, 213003 (2013).

\bibitem{Schwarzkopf2011}
A. Schwarzkopf, R.E. Sapiro, and G. Raithel,
Phys. Rev. Lett. {\bf 107}, 103001 (2011).

\bibitem{Viteau2011}
M. Viteau, M.G. Bason, J. Radogostowicz, N. Malossi, D. Ciampini, 
O. Morsch, and E. Arimondo,
Phys. Rev. Lett. {\bf 107}, 060402 (2011).

\bibitem{Schauss2012}
P. Schau\ss, M. Cheneau, M. Endres, T. Fukuhara, S. Hild, A. Omran,
T. Pohl, C. Gross, S. Kuhr, and I. Bloch,	
Nature {\bf 491}, 87 (2012).

\bibitem{Raithel2005}
T. Cubel Liebisch, A. Reinhard, P.R. Berman, and G. Raithel,
Phys. Rev. Lett. {\bf 95}, 253002 (2005).

\bibitem{Viteau2012}
M. Viteau, P. Huillery, M.G. Bason, N. Malossi, D. Ciampini, 
O. Morsch, E. Arimondo, D. Comparat, and P. Pillet,
Phys. Rev. Lett. {\bf 109}, 053002 (2012).

\bibitem{Hofmann2013}
C.S. Hofmann, G. G\"unter, H. Schempp, M. Robert-de-Saint-Vincent, 
M. G\"arttner, J. Evers, S. Whitlock, and M. Weidem\"uller,
Phys. Rev. Lett. {\bf 110}, 203601 (2013)

\bibitem{Weimer2008}
H. Weimer, R. L\"ow, T. Pfau, and H.P. B\"uchler,
Phys. Rev. Lett. {\bf 101}, 250601 (2008).

\bibitem{Weimer10}
H. Weimer and H.P. B\"uchler, 
Phys. Rev. Lett. {\bf 105}, 230403 (2010).

\bibitem{Pohl2010}
T. Pohl, E. Demler, and M.D. Lukin,
Phys. Rev. Lett. {\bf 104}, 043002 (2010).

\bibitem{Schachenmayer2010}
J. Schachenmayer, I. Lesanovsky, A. Micheli, and A.J. Daley,
New J. Phys. {\bf 12}, 103044 (2010).

\bibitem{Bijnen2011}
R.M.W. van Bijnen, S. Smit, K.A.H. van Leeuwen, 
E.J.D. Vredenbregt, and S.J.J.M.F. Kokkelmans,
J. Phys. B {\bf 44}, 184008 (2011).

\bibitem{Sela2011}
E. Sela, M. Punk, and M. Garst,
Phys. Rev. B {\bf 84}, 085434 (2011).

\bibitem{Lesanovsky2011}
I. Lesanovsky,
Phys. Rev. Lett. {\bf 106}, 025301 (2011).

\bibitem{Lesanovsky2012}
I. Lesanovsky,
Phys. Rev. Lett. {\bf 108}, 105301 (2012).

\bibitem{Garttner2012}
M. G\"arttner, K.P. Heeg, T. Gasenzer and J. Evers,
Phys. Rev. A {\bf 86}, 033422 (2012).

\bibitem{Zeller2012}
W. Zeller, M. Mayle, T. Bonato, G. Reinelt, and P. Schmelcher,
Phys. Rev. A {\bf 85}, 063603 (2012). 

\bibitem{Lee2011}
T.E. Lee, H. H\"affner, and M.C. Cross,
Phys. Rev. A {\bf 84}, 031402(R) (2011).

\bibitem{Qian2012}
J. Qian, G. Dong, L. Zhou, and W. Zhang,
Phys. Rev. A {\bf 85}, 065401 (2012).

\bibitem{Hoening2013}
M. H\"oning, D. Muth, D. Petrosyan, and M. Fleischhauer, 
Phys. Rev. A {\bf 87}, 023401 (2013).

\bibitem{Ji2011}
S. Ji, C. Ates and I. Lesanovsky,
Phys. Rev. Lett. {\bf 107}, 060406 (2011).

\bibitem{AtesGarraha2012}
C. Ates, J. P. Garrahan and I. Lesanovsky
Phys. Rev. Lett. {\bf 108}, 110603 (2012).

\bibitem{Ates2012Ji2013}
C. Ates and I. Lesanovsky,
Phys. Rev. A {\bf 86}, 013408 (2012);
S. Ji, C. Ates, J.P. Garrahan and I. Lesanovsky,
J. Stat. Mech. P02005 (2013).

\bibitem{DPMHMF2013}
D. Petrosyan, M. H\"oning, and M. Fleischhauer, 
Phys. Rev. A {\bf 87}, 053414 (2013).

\bibitem{Petrosyan2013}
D. Petrosyan, 
J. Phys. B {\bf 46}, 141001 (2013).

\bibitem{Ates2007}
C. Ates, T. Pohl, T. Pattard, and J.M. Rost, 
Phys. Rev. A {\bf 76}, 013413 (2007).

\bibitem{Ates2011}
C. Ates, S. Sevin\c{c}li, and T. Pohl,
Phys. Rev. A {\bf 83}, 041802(R) (2011).

\bibitem{Heeg2012}
K.P. Heeg, M. G\"arttner and J. Evers,
Phys. Rev. A {\bf 86}, 063421 (2012).

\bibitem{PLDP2007}
P. Lambropoulos and D. Petrosyan,
{\it Fundamentals of Quantum Optics and Quantum Information},
(Springer, Berlin, 2007).

\bibitem{rydcalc}
K. Singer, J. Stanojevic, M. Weidem\"uller, and R. C\^ot\'e,
J. Phys. B {\bf 38}, S295 (2005).

\bibitem{MandelQ}
L. Mandel, 
Opt. Lett. {\bf 4}, 205 (1979).


\end{thebibliography}
\end{document}